\documentstyle[prl,aps,floats,epsf,color]{revtex}
\begin{document}
\baselineskip=12pt
\def\be{\begin{equation}}
\def\ee{\end{equation}}
\def\bea{\begin{eqnarray}}
\def\eea{\end{eqnarray}}
\def\E{{\rm e}}
\def\bearst{\begin{eqnarray*}}
\def\eearst{\end{eqnarray*}}
\def\peleven{\parbox{11cm}}
\def\peffec{\peight{\bearst\eearst}\hfill\peleven}
\def\pspace{\peight{\bearst\eearst}\hfill}
\def\ptwelve{\parbox{12cm}}
\def\peight{\parbox{8mm}}
\twocolumn[\hsize\textwidth\columnwidth\hsize\csname@twocolumnfalse\endcsname

\title
{Stochastic Analysis and Regeneration of Rough Surfaces}

\author
{ G. R. Jafari $^a$, S. M. Fazeli $^a$, F. Ghasemi $^a$, S. M.
Vaez Allaei $^b$, \\ M. Reza Rahimi Tabar $^{a,c,d}$, A. Iraji zad
$^a$  and G. Kavei $^e$}
\address
{ $^a$  Department of Physics, Sharif University of
Technology, P.O. Box 11365-9161, Tehran, Iran \\
$^b$ Institute for Advanced Studies in Basic Sciences, P. O. Box
45195-159, Zanjan  Iran \\
$^c$ CNRS UMR 6529, Observatoire de la C$\hat o$te d'Azur,
BP 4229, 06304 Nice Cedex 4, France\\
$^d$ Dept. of Physics, Iran University of Science and Technology,
Narmak, Tehran 16844, Iran\\
$^e$ Material and Energy, Research Center, P. O. Box 14155-4777,
Tehran, Iran \\}
 \maketitle


\begin{abstract}

We investigate Markov property of rough surfaces. Using
stochastic analysis we characterize the complexity of the surface
   roughness by means of a Fokker-Planck or Langevin equation.
   The obtained Langevin equation enables us to regenerate
   surfaces with similar statistical properties compared with the
   observed morphology by atomic force microscopy.

\end{abstract}
\hspace{.3in}
\newpage
] Studying the growth, formation and morphology of interfaces has
been one of the recent interesting fields of study because of its
high technical and rich theoretical advantages [1].
 One of the main problems in this area is the scaling
behaviour of the moments of height difference $ \Delta h = h (x_1)
- h (x_2) $ and the evolution of the probability density function
(PDF) of $ \Delta h $, i.e. $P(\Delta h, \Delta x)$ in terms of
the length scale $\Delta x$.
 Recently Friedrich and Peinke
have been able to obtain a Fokker--Planck equation describing the
evolution of the probability distribution function in terms of the
length scale, by analyzing some stochastic phenomena, such as
turbulent free jet, etc. [2-4]. They noticed that the conditional
probability density of field increments (velocity field, etc.)
satisfies the Chapman-Kolmogorov equation. Mathematically this is
a necessary condition for the fluctuating data to be a Markovian
process in the length scales [5].

In this letter using the method proposed by Friedrich and Peinke,
we measure the Kramers--Moyal`s (KM) coefficients for the
fluctuating fields $\Delta h$ and $h(x)$ of a deposited copper
film. It is shown that the first and second KM`s coefficients
have well--defined values, while the third and fourth order
coefficients tend to zero. Therefore, by addressing the
implications dictated by the theorem [5] a Fokker--Planck
evolution operator has been found. The Fokker--Planck equation for
$P(\Delta h, \Delta x)$ is used to give information on changing
the shape of PDF as a function of the length scale $ \Delta x$.
By using this strategy the information of the observed
intermittency of the height fluctuation is verified [6]. The first
and second KM`s coefficients for the fluctuations of $h(x)$,
enables us to write a Langevin equation for the evolution of
height with respect to $x$. Using this equation we regenerate the
surface with similar statistical properties, compared with the
   observed morphology by atomic force microscopy. The regeneration of a surface is known
as the inverse method. There are other inverse method approaches
introduced in the literature [13]. In the previous attempts, to
regenerate the surface, an evolution equation for $h(x,t)$ vs $t$
has been evaluated. Here we do this by an evolution equation for
$h(x)$ vs $x$, for a certain time.

For this purpose, a copper film was deposited on a polished
Si(100) substrate by the resistive evaporation method in a high
vacuum chamber. The pressure during evaporation was $10^{-6}$
Torr. The thickness of the growing films was measured in situ by
a quartz crystal thickness monitor. We performed all depositions
at room temperature, with a deposition rate about $20-30$$
nm/min$. The substrate temperature was determined using a
chromel/alumel thermocouple mounted in close proximity of
samples. The surface topography of the films was investigated
using Park Scientific Instruments model Autoprobe CP. The images
were collected in a constant force mode and digitized into $256
\times 256 $ pixels with scanning frequency of $0.6$ Hz. The
cantilever of 0.05 N $m^{-1}$ spring constant with a commercial
standard pyramidal $Si_3N_4$ tips was used. A variety of scans,
each with size $L$ were recorded at random locations on the $Cu$
film surface.

It is a common procedure to characterize the complexity of a rough
surface by checking the scaling behaviour of the moments $C_q =
<|h(x_1) - h(x_2)|^q >$ in terms of the length scale $\Delta x =
|x_1 - x_2|$. We investigated the scaling behaviour of the q-th
moment $C_q$ and observed that all of the moments (up to $q=20$)
behave as $|x_1 - x_2|^{\xi_q}$ within the scaling region $\sim
10 $ to $150$ $nm$. We have found a nonlinear relation between
$\xi_q$ and $q$. This shows that the height fluctuations are
intermittent or multi--fractal ( see [8, 15 ] and references
therein). The roughness exponent $\alpha$ is related to the
exponent $\xi_2$ as $\alpha=\xi_2/2$ [1]. For the stationary
samples with thickness $440$ $nm$, the roughness exponent
$\alpha$ was found to be  $ 0.83
 \pm 0.03$. From the stochastic point of view one has to remark that
multi-fractality is based on properties of the roughness on
distinct length scales. However, checking the scaling behaviour
does not explain possible correlation between the roughness
measures on different scales.
 Also it is noted that the methods based on multi-fractality are limited to
the subclass of rough surfaces which show scaling properties. The
method introduced by Friedrich and Peinke is a general method,
which explains the complexity of the surface roughness, with no
scaling feature to be explicitly required.
 Their method yields an estimation
of an effective stochastic equation in the form of a Fokker-Planck
equation (also known as Kolmogorov equation). The connection
between the multifractality and Markovianity has been discussed
in [4].

\begin{figure}
\epsfxsize=6.5truecm\epsfbox{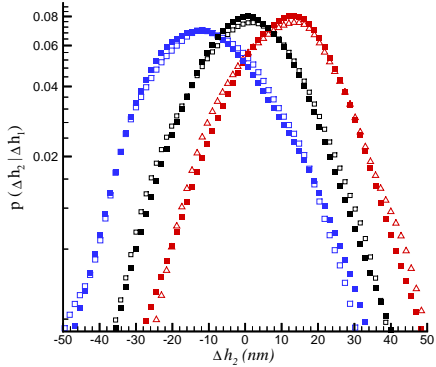}
 \narrowtext \caption{ Test of Chapman-Kolmogorov equation for
 different values $\Delta h_1 = - 21nm $, $\Delta h_1 = 0 $  and  $\Delta h_1 = 21nm$.
  The bold and open symbols represent
 directly evaluated PDF and the integrated PDF, respectively.
The length scales $\Delta x_1$, $\Delta x_2$ and $\Delta x_3$ are
$180nm$, $320nm$ and $260 nm$, respectively.}
 \end{figure}


A complete characterization of the statistical properties of the
height fluctuation requires the evaluation of joint PDF`s
$P_N(\Delta h_1,\Delta x_1;....;\Delta h_N,\Delta x_N)$, for any
arbitrarily $N$. If the process is a Markov process (a process
without memory), an important simplification arises. For this
type process the N-point joint PDF,  $P_N$, is generated by a
product of the conditional probabilities $P(\Delta h_{i+1},\Delta
x_{i+1}|\Delta h_i,\Delta x_i)$, for $i=1,...,N-1$. As a
necessary condition for being a Markov process, the
Chapman-Kolmogorov equation,

\bea
  &&p(\Delta h_2,\Delta x_2|\Delta h_1,\Delta x_1)= \cr \nonumber\\
  &&\int \hbox{d} (\Delta h_3)\,
  p(\Delta h_2,\Delta x_2|\Delta h_3,\Delta x_3)\, p(\Delta h_3,\Delta
  x_3|\Delta h_1,\Delta x_1)
\eea

 should hold for any value of $\Delta x_3$, in the
interval $  \Delta x_2 < \Delta x_3 < \Delta x_1 $ [5].
 We checked
the validity of the Chapman-Kolmogorov equation for different
$\Delta h_1$ triplets by comparing the directly evaluated
conditional probability distributions $p(\Delta h_2,\Delta
x_2|\Delta h_1,\Delta x_1)$ with the ones calculated according to
rhs. of eq.(1). In Fig. (1), the two direct and integrated PDF`s
 are superimposed for the purpose of illustration. The bold and
open symbols represent directly evaluated PDF and the integrated
PDF, respectively. Assuming a statistical error of the square
root of the number of events of each bin we find that both PDF`s
are statistically identical (see also [14] an another interesting
and carefully presented example of
   application of the Chapman-Kolmogorov equation).

 It is well-known, the Chapman-Kolmogorov
equation yields an evolution equation for the change of the
distribution function $p(\Delta h, \Delta x)$ across the scales
$\Delta x$. The Chapman-Kolmogorov equation formulated in
differential form yields a master equation, which can take the
form of a Fokker-Planck equation [5]:
 \bea
  &&\frac {d}{d r} p(\Delta h,r)=\cr \nonumber\\
 && [-\frac{\partial }{\partial \Delta h}
  D^{(1)}(\Delta h, r)
  +\frac{\partial^2 }{\partial \Delta h^2} D^{(2)}(\Delta h, r)]
  p(\Delta h, r)
\eea

\begin{figure}
\epsfxsize=6.5truecm\epsfbox{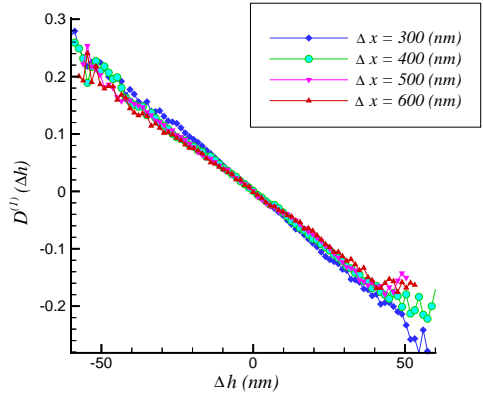}
\epsfxsize=6.5truecm\epsfbox{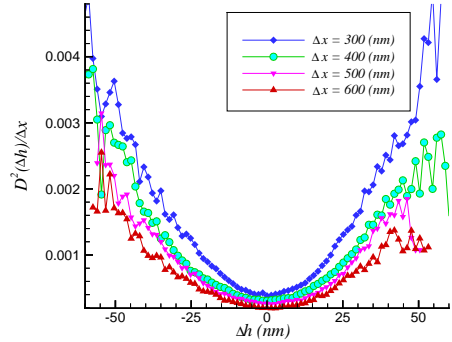} \narrowtext \caption{ Drift
and diffusion coefficients $D^1 ( \Delta h )$ and $D^2( \Delta h)$
are estimated from the eq. (3). The $D^1$ and $D^2$ present the
linear and quadratic behavior, respectively.}
\end{figure}

\begin{figure}
\epsfxsize=6.0truecm\epsfbox{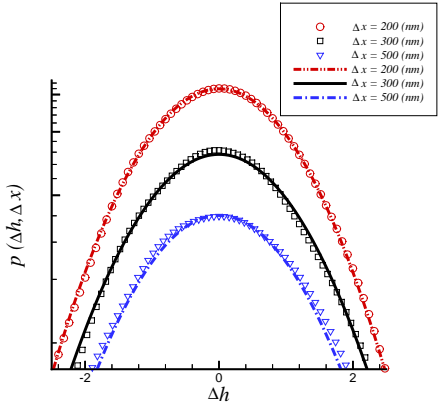}
 \narrowtext \caption{Probability densities of the height difference
 $\Delta h = h (x + \Delta x) - h(x)$ for the length scales $\Delta x = 200, 300$ and $500 $ $nm $
(from top to bottom).  The results obtains from the data analysis
of the AFM image (R) and numerical integration of an effective
Fokker-Planck equation (E) i.e. eq.(2), respectively.
 The PDF`s are shifted in vertical directions for convenience of
presentation and $\Delta h$`s are measured in units of the
standard deviation of $\Delta h$ at $\Delta x = 200 nm$.
 }
\end{figure}

 where $r:=\Delta x$. The drift and diffusion coefficients $D^{(1)}(\Delta h,
r)$, $D^{(2)}(\Delta h, r)$ can be estimated directly from the
data and the moments $M^{(k)}$ of the conditional probability
distributions:
 \bea
  && D^{(k)}(\Delta h,r) = \frac{1}{k!}
   \hskip .2cm lim_{\Delta r \rightarrow 0}  M^{(k)} \cr \nonumber \\
  && M^{(k)} = \frac{1}{\Delta r}  \int d\Delta h'
  (\Delta h'-\Delta h)^k p(\Delta h', r+\Delta r|\Delta h, r) .
 \eea

The coefficients $D^{(k)}(\Delta h,r)$`s are known as Kramers-Moyal coefficients.
The drift and diffusion coefficients $ D^{(1)}$ and $D^{(2)}$ are
displayed in fig.(2). It turns out that the drift term $D^{(1)}$
is a linear function of $\Delta h$, whereas the diffusion term
$D^{(2)}$ is a function quadratic in $\Delta h$. For large values
of $\Delta h$, our estimation becomes poor and thus uncertainty
increases. From the analysis of the data set we obtain the
following approximation:

\bea
&& D^{(1)} ( \Delta h, \Delta x) =  - 0.0055  \hskip .2 cm \Delta h  \cr \nonumber \\
&&D^{(2)} (\Delta h, \Delta x) = [(2.9 \times 10^{-4}) (\Delta
h)^2 + 0.015 (\Delta x)^{0.45}]/{\Delta x} \eea

 where $\Delta h$ is measured in units of the standard deviation of
$\Delta h$ at $\Delta x = 200 nm$.
According to Pawula`s theorem,
the Kramers-Moyal expansion stops after the second term, provided
that the fourth order coefficient $D^{(4)} ( \Delta h, \Delta x )$
vanishes [5].
 The forth order coefficients $D^{(4)}$
 in our analysis was found to be about $ {D^{(4)}} \simeq 10^{-4} {D^{(2)}}$. In this
approximation we can ignore the coefficients $D^{(n)}$ for $n \geq
3$. To perform a quantitative test of the result with these
coefficients, we solve the Fokker-Planck equation for the PDF at
scales $ \Delta x \ll L $ with a given distribution at sample
size $L$ [6,7]. Fig.(3) shows a comparison between the analysis
of AFM image and the solutions of the obtained Fokker-Planck
equation for the copper surface for the length scales $\Delta x =
200, 300$ and $500$ $nm$. The figure shows that the solutions of
our model fit the experimentally determined PDF's with good
precision. In the integral scale our measured PDF is nearly a
Gaussian distribution. In our approximation the stochastic process
underlying the height fluctuation changes is a linear stochastic
process with multiplicative noise.

 By the same procedure, we checked
the Markovian nature of the fluctuations of the height $ h =
h(x)-\bar h $, and found the following expression for the
$D^{(1)} (h)$ and $D^{(2)}(h)$:

\bea
&& D^{(1)} (h) = - 0.01 h  \cr \nonumber \\
&& D^{(2)}(h) = 0.088 - 0.004 h  + 5.19 \times 10^{-5}  h^2. \eea

 The height field is measured in units of the standard deviation
of $h$. Analogous to equation (2), we can write a Fokker-Planck
equation for the PDF of $h$ by replacing $ r $ and $\Delta h$
with $x$ and $h$, respectively.
 We note that this
Fokker-Planck equation is equivalent to the following  Langevin
equation (using the Ito interpretation) [5]:

\begin{equation}
  \frac{d}{d x}  h(x)=D^{(1)}( h ) +
  \sqrt{D^{(2)}( h )} \hskip .1 cm f(x) \qquad .
\end{equation}

Here, $f(x)$ is a random force, zero mean with gaussian
statistics, $\delta$-correlated in $x$, i.e.
$<f(x)f(x')>=\delta(x-x')$. Furthermore, with this last
expression, it becomes clear that we are able to separate the
deterministic and the noisy components of the surface height
fluctuations in terms of the coefficients $D^{(1)}$ and
$D^{(2)}$. Equation (6) enables us to regenerate rough surfaces
which are similar to the original one (in the statistical sense
). In fig. (4) the AFM  and  regenerated images are demonstrated.
The regenerated surface is very similar in statistical sense to
the original one. All regenerated patterns are statistically
similar. To ensure this fact, for instance, in Fig.(5) we have
plotted the second moment of the structure function $C_2$  for
the AFM and a regenerated surfaces and their roughness exponents
were found  $0.83\pm 0.03$ and $ 0.83 \pm 0.01$, respectively.
\begin{figure}
\epsfxsize=7.5truecm\epsfbox{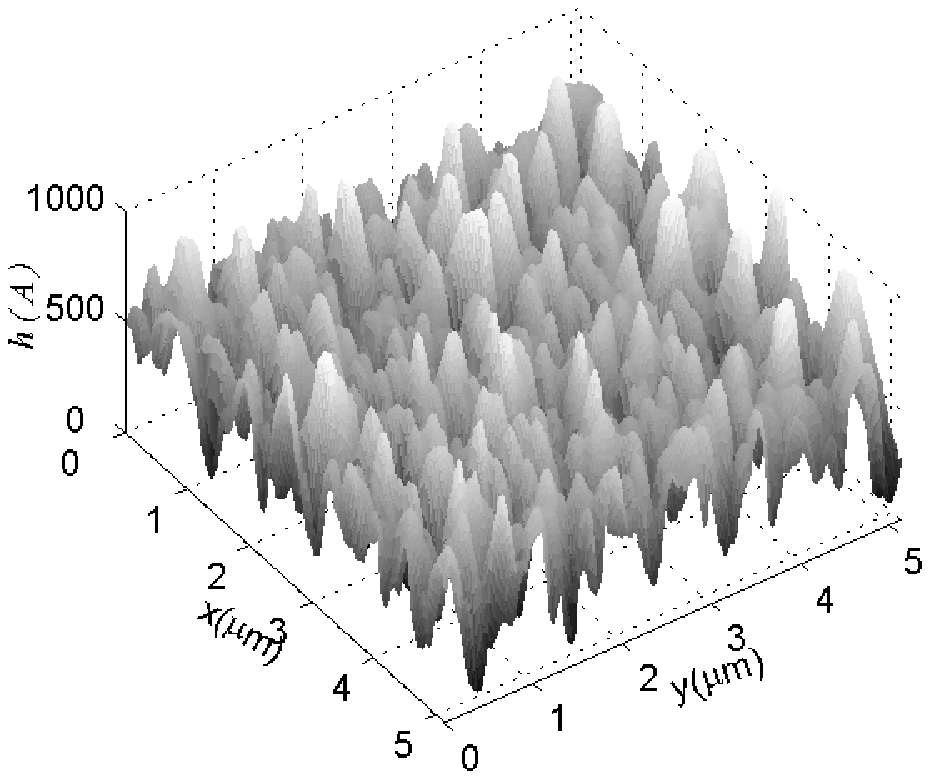}
\epsfxsize=7.9truecm\epsfbox{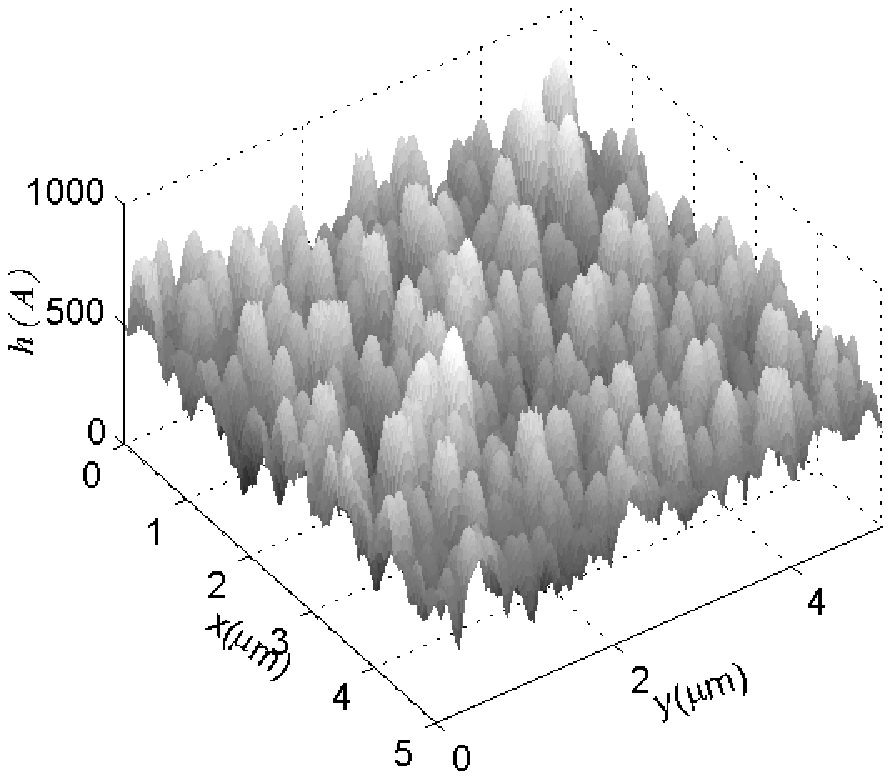}
 \narrowtext \caption{ AFM  and regenerated surface images (from up to bottom), which we
have regenerated the rough surface using the Langevin equation for
dynamics of h(x). As drift term $ D^{(1)} (h) = - 0.01 h $ and as
diffusion term $D^{(2)}(h) = 0.088 - 0.004 h  + 5.19 \times
10^{-5} h^2 $.}
\end{figure}
\begin{figure}
 \epsfxsize=6.0truecm\epsfbox{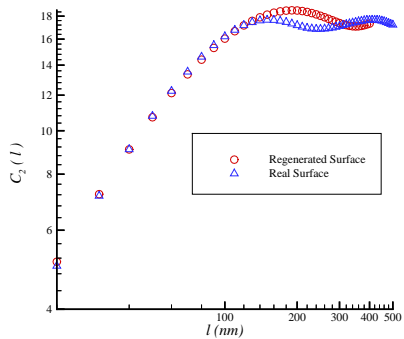}
 \narrowtext \caption{Log-log plot of the second moment of height-difference vs $l$,
for real sample and regenerated sample. The roughness exponents
for  real and regenerated are $0.83\pm 0.03$ and $ 0.83 \pm
0.01$, respectively. }
\end{figure}


There are a few comments on the regeneration of rough surface
that we would like to notify.
 When we are discussing about a Markov process, one should note
 that, this is true within an approximation. For instance, in random motion of a particle inside a fluid,
  it is known that
  the collision
of the particle with the fluid molecules is not instantaneous, and
takes a certain duration. During the time that a collision is
taking place, the change of velocity is not Markov, because the
velocities in the collision time scale have memory.
Consequently,  in the time series for the velocities of the
particle, if the time intervals are less than the collision time
scale, the process can not be regarded as a Markov process. The
minimum time interval that the particle motion can be considered
as a Markov process is known as Markov time scale and the motion
is known as a Brownian motion.
 In the stochastic analysis of the rough surface, we are dealing
with the Markov property of height fluctuations in spatial
dimensions, therefore, instead of a Markov time scale, here we
will have a Markov length scale $l_{markov}$.
 Our analysis shows that $l_{markov} \simeq  160$ $nm $,
  equivalent to 8 pixels in our AFM image [9].
The surface is regenerated by iterating the eq.(6), which gives us
a series of data without memory. To compare the regenerated
surface with the original one, we have to take the spatial
interval in the numerical discritization of eq.(6), to be equal
to 1 pixel. However, here the Markov length is equal to $8$
pixels.
 Therefore, we should relate the height
field within the Markov length. There are a number of methods to
correlate the generated data in this interval [9]. We do this by
means of scanning the surface with a tip, where its size is about
the Markov length [12]. The tip that we have used has the form $z
= a x^2 + b y^2$. In this case the parameters $a$ and $b$ are
$0.035$.

In summary, we have shown that the probability density of height
increments satisfy a Fokker-Planck equation which encodes the
Markovian property of these fluctuations in a necessary way. We
are able to give the expression of Kramers-Moyal`s coefficients
for the stochastic processes $\Delta h$ and $h$ by using the
polynomial ansatz [10,11]. Also we could find the form of path
probability functional of the height increments in spatial scale,
which naturally encodes the scale dependence of probability
density. This gives a clear picture about the intermittent nature
in height fluctuations. The methods enables us to regenerate many
realizations of the rough surface with similar statistical
properties in favored scales. As an application, large surface
generation would be possible by sampling the real surface with
high resolution (in same the resolution as nanoscope imaging, e.g.
AFM images). This would be applicable in computer simulation of
the surface and interface processes, for example, the diffusion of
materials between rough surfaces, the effect of roughness on the
friction, and so on.

We thank  F. Azami, M. Sahimi, and  N. Taghavinia for their useful
discussions.


\begin{thebibliography}{99}


\bibitem{1}J. Krug and H. Spohn in
"Solids Far from Equilibrium Growth, Morphology and Defects",
 edited by C. Godreche (Cambridge University Press, New York,
 1990); A.-L. Barabasi and H. E. Stanley, "Fractal Concepts in Surface Growth"
 (Cambridge University Press, New York, 1995) ; T. Halpin-Healy and Y. C. Zhang, Phys. Rep.{\bf
 245},218(1995).
\bibitem{2} R. Friedrich and J. Peinke Phys. Rev. Lett. {\bf 78},
863 (1997)
\bibitem{3} R. Friedrich, J. Peinke and C. Renner Phys. Rev. Lett. {\bf
84}(22),5224 (2000)
\bibitem{4} R. Friedrich, K.Marzinzik and A.Schmigel, in "A perspective Look
at Nonlinear Media", Edited by Jurgen Parisi, Stefan C. Muller and
Walter Zimmermann, Lecture notes in Physics, Vol. {\bf 503}, P.
313 (Springer-verlag, Berlin, 1997).;R. Friedrich, S. Siegert, J.
Peinke, etal.  Physics Letters A,  {\bf271}(3):217, (2000).
\bibitem{5} H. Risken "The Fokker-Planck equation" Springer, Berlin, 1984.
\bibitem{6} J. Davoudi and M. Reza Rahimi Tabar, Phys. Rev. Lett. {\bf
82}, 1680 (1999)
\bibitem{7} A. A. Donkov, A.D. Donkov and E.I. Grancharova,
math-ph/9807010, math-phys/9807011
\bibitem{8} A. Iraji zad, G. Kavei, M. Reza Rahimi Tabar and S.M.
Vaez Allaei, J. Phys. : Condensed Matter {\bf 15} 1889 (2003)
\bibitem{9} C. Renner, J. Peinke, and R. Friedrich.
Journal of Fluid Mechanics, {\bf 433} 383, (2001);
cond-mat/0102494.
\bibitem{10} M. Ragwitz and H. Kantz, Phys. Rev. Lett. {\bf87}, 254501 (2001)
\bibitem{11} R. Friedrich, C. Renner, M.Siefert and J. Peinke,
Physics/0203005
\bibitem{12} J. Aue and J. Th. M. De Hosson, Appl. Phys. Lett 71
(10) 1347 (1997)
\bibitem{13} C-H.  Lam and L.M. Sander, Phys. Rev. Lett {\bf 71}, 561 (1993);
A. Giacometti and M. Rossi, Phys. Pev. E {\bf 63} 046102 (2001)
\bibitem{14} A. Fuliski, Z. Grzywna, I. Mellor, Z. Siwy, and P. N. R.
   Usherwood, Phys. Rev. E., {\bf 58} 919, 1998.
\bibitem{15} S. Mercik and  K. Weron, Phys. Rev. E  {\bf 63} 051910, 2001

\end{thebibliography}
\end{document}